\documentclass[pra,twocolumn,showpacs,preprintnumbers,amsmath,amssymb]{revtex4}

\usepackage{epsf}
\usepackage{amssymb}

\newcommand{\ket}[1]{\mathop{\left| #1 \right\rangle}\nolimits}
\newcommand{\bra}[1]{\mathop{\left\langle #1 \right|}\nolimits}
\newcommand{\braket}[2]{\mathop{\left\langle #1 \left| #2 \right.
\right\rangle}\nolimits}

\begin{document}

\title{Entangling quantum measurement and its properties}

\author{Boris A. Grishanin}
 \email{grishan@comsim1.phys.msu.su}
\author{Victor N. Zadkov}
 \email{zadkov@comsim1.phys.msu.su}
\affiliation{International Laser Center and Department of Physics\\
M.\ V.\ Lomonosov Moscow State University, 119899 Moscow, Russia}

\date{April 14, 2003}

\begin{abstract}
We study the mathematical structure of superoperators describing quantum
measurements, including the \emph{entangling measurement}---the
generalization of the standard quantum measurement that results in
entanglement between the measurable system and apparatus.
It is shown that the coherent information can be effectively used
for the analysis of such entangling measurements whose possible
applications are discussed as well.
\end{abstract}


\maketitle

\section{\label{section:intro}Introduction}

Experiments in the field of quantum information processing and
engineering, a new emerging interdisciplinary field of science
\cite{QC}, require the acquisition of information about the quantum
system ($A$, hereafter, the ``object") by means of an apparatus that
produces a \emph{measurement} \cite{Neumann55,zurek,Ludwig}.
During the measurement, the object quantum system and apparatus
interact with each other (and with the environment or the ``reference"
system).  As a result, the apparatus acquires essentially quantum
information initially contained in the measurable quantum system.

The measurement procedure and the structure of the related measurement
transformation may vary essentially. For example, in quantum optics the
\emph{coherent} measurement transformation is used
\cite{gisin01,cohmeas}. In applications to the emerging field of quantum
information processing and quantum computing, quantum measurement can
also be considered as an effective tool for realizing quantum
algorithms \cite{childs02}. Although quantum measurement procedures
can vary significantly for different applications it is worth selecting
and examining mathematical forms of common types of measurement, their
properties, and areas of possible applications.

In this work, we examine a class of quantum measurements
completely preserving the initial concept of quantum measurement as the
wave function collapse, $\psi\to\{p_n=|c_n|^2, \ket n\}$
\cite{sudbery,Measur}, i.e., the transfer of an initially pure state into a
mixed ensemble of pure orthogonal states $\ket n$ with probabilities
$p_n$. We call such a measurement a \emph{standard} measurement.
The initial coherency in the object system (in the initial wave function
$\psi=\sum c_n\ket n$ between the eigenstates of the measurable variable
$\hat A=\sum\lambda_n\ket n\bra n$) is therefore completely lost.
Since it happens for \emph{any} quantum state $\psi$, this means that
the object quantum system is completely \emph{dequantized}, i.e., the
only subset of orthogonal states $\ket n$ out of all the system states
$\psi$ is left. This subset is equivalent to a set of classical events.
The final states of the apparatus can then be characterized by a measurable
variable $\hat M$, which we will call a ``pointer" \cite{preskill}.
After the measurement, the pointer has the same values as those of
$\hat A$.

The generalization of this concept discussed here lies in considering
a more general set of states \emph{after} the quantum measurement,
which takes into account the created entanglement in the
object--pointer system after the measurement. We will call such a
measurement an \emph{entangling} measurement. Note that the
entangling measurement introduced here is qualitatively different
from the measurement transformation defined in Ref.\
\onlinecite{vedral02}, where it is introduced in the form of a
unitary transformation, and from the transformation defined in Refs.\
\onlinecite{ozawa01,barnum02}, where the quantum properties of states
in the bipartite system object--pointer are not considered. Also,
it is worth noting that the measurement transformations in the
bipartite system are clearly related to the
characterization of transformations in the bipartite setting
Alice--Bob restricted by physical causality relations
\cite{Preskill-pap}.

One of the fundamental properties of a measurement transformation is that
the resulting state of the bipartite system object--pointer does not
depend on the initial state of the apparatus. In case of an entangling
measurement, this property, which makes the transformation irreversible,
distinguishes it from reversible transformations of quantum entanglement,
which play a fundamental role in quantum information processing \cite{QC}
and can be realized with the help of a unitary transformation applied to the
bipartite system.

In quantum information theory, entanglement is one of the key concepts
used for characterization of quantum states of a bipartite setting. It
determines the quantum specifics of physical interaction on which such
its practical applications as quantum computing and quantum cryptography
rely \cite{QC,UFN01}. For a bipartite system object--pointer in a pure state
described by a joint wave function $\psi_{AM}$, there is a unique
definition of the degree of quantum entanglement as the entropy of
separate density matrices $S(\hat\rho^A)=S(\hat\rho^M)$, where
$\hat\rho^A={\rm Tr}_M\psi_{AM} \psi_{AM}^+$ and $\hat\rho^M ={\rm Tr}_A
\psi_{AM} \psi_{AM}^+$. However, for the case of mixed states there is
no unique valid definition of the degree of entanglement.

In this connection, it is worth noting that the entanglement of
quantum states created here is due to the above-specified
alternative choice of the quantum measurement procedure. One can
even assume that the characterization of the degree of entanglement
as a derivative from the transformation measurement structure
realized by an apparatus can give some additional information about
the physical contents of the entanglement concept.

In Sec.\ \ref{section:definitions}, we give precise mathematical
definitions of both standard and entangling measurement transformations,
special cases of which are considered in Sec.\ \ref{section:cases}. We
describe the general structure of a quantum measurement superoperator and
specify structure of the entanglement matrix and the entangling
measurement superoperator in Sec.\ \ref{section:matrix}. The discussed
formalism is clarified by an example of a two-dimensional model. In Sec.\
\ref{section:exchange} we address the question whether an entangling measurement can be
used for an entanglement transfer in various applications of quantum
information processing. We prove that it cannot be used for the entanglement
transfer from a bipartite system to another one. We consider the quantitative
characteristics of entanglement in Sec.\ \ref{section:entangl} arguing that
the coherent information is a valid tool for characterizing the
entanglement created in the bipartite system object--pointer.

\section{\label{section:definitions}Mathematical definitions of quantum measurement transformations}

Following the traditional quantum measurement postulate
\cite{Neumann55,zurek,Ludwig,Measur}, the process of quantum measurement
of a quantum object system $A$ that lives in Hilbert space $H_A$ by the
classical pointer variable $M$ leads to establishing the resulting state
of the system $A$. In this state the measurable physical variable described
by a quantum operator $\hat A$ takes one of the range of possible values $\lambda$
and the apparatus' classical variable $M$ coincides with that value: $M=\lambda$.
We then can simplify the description of the apparatus preserving only the initial
($\mu$) and resulting ($\lambda$) values of the pointer variable. After
this simplification, the measurement transformation in the bipartite
system ``quantum object--pointer'' is represented by the superoperator quantum
transformation along the quantum variables of the measurable system $A$ and
the classical conditional probability distribution along the classical pointer
variables $\mu$ of the apparatus \cite{QSP}:
\begin{equation}\label{Slm}
{\cal M}(\lambda|\mu)=\hat P_\lambda^A\odot\hat P_\lambda^A.
\end{equation}

A symbolic representation of the quantum state transformation operators
is used here, in which the substitution symbol $\odot$ is to be
substituted by a transformed operator. The operators $\hat P_\lambda^A$
are the orthogonal projectors onto the subspaces with eigenvalue $\lambda$ of the
measurable variable $\hat A=\sum\lambda\hat P_\lambda^A$, where $\lambda$
enumerates the final states of the apparatus (there is no dependence on $\mu$,
which indicates independence from the initial state of the pointer).

The initial states of the bipartite system ``quantum object
system--apparatus" are described by the joint quantum-classical
distributions $\hat\rho(\mu)$ defined as the linear operators in
the direct product $H_A\otimes \Lambda_M$ of the Hilbert space of
the quantum system and the set $\Lambda_M$ of classical values of
the pointer variable. They obey the positivity requirement
$\hat\rho(\mu)\ge0$ (i.e., $\bra\psi \hat\rho (\mu)\ket\psi\geqq0$
for any $\psi$ and all $\mu$) and the normalization condition
$\sum_\mu{\rm Tr}\,\hat\rho(\mu)=1$ and are transformed with the
help of the superoperator (\ref{Slm}) as
\begin{equation}\label{rlm}
\hat\rho(\mu)\to\hat\rho(\lambda)=\sum_\mu{\cal M}(\lambda|\mu)\hat
\rho(\mu)=\sum_\mu\hat P_\lambda^A\hat\rho(\mu)\hat P_\lambda^A.
\end{equation}

Here we restrict our consideration to the class of ``direct" measurements
that are described by the orthogonal projectors $\hat P_\lambda^A$,
having in mind the fundamental character of this subclass of all possible
measurements. More general measurements can then be smoothly handled in
the open systems framework.

\begin{figure}[th]
\begin{center}
 \epsfxsize=0.25\textwidth \epsfclipon  \leavevmode \epsffile{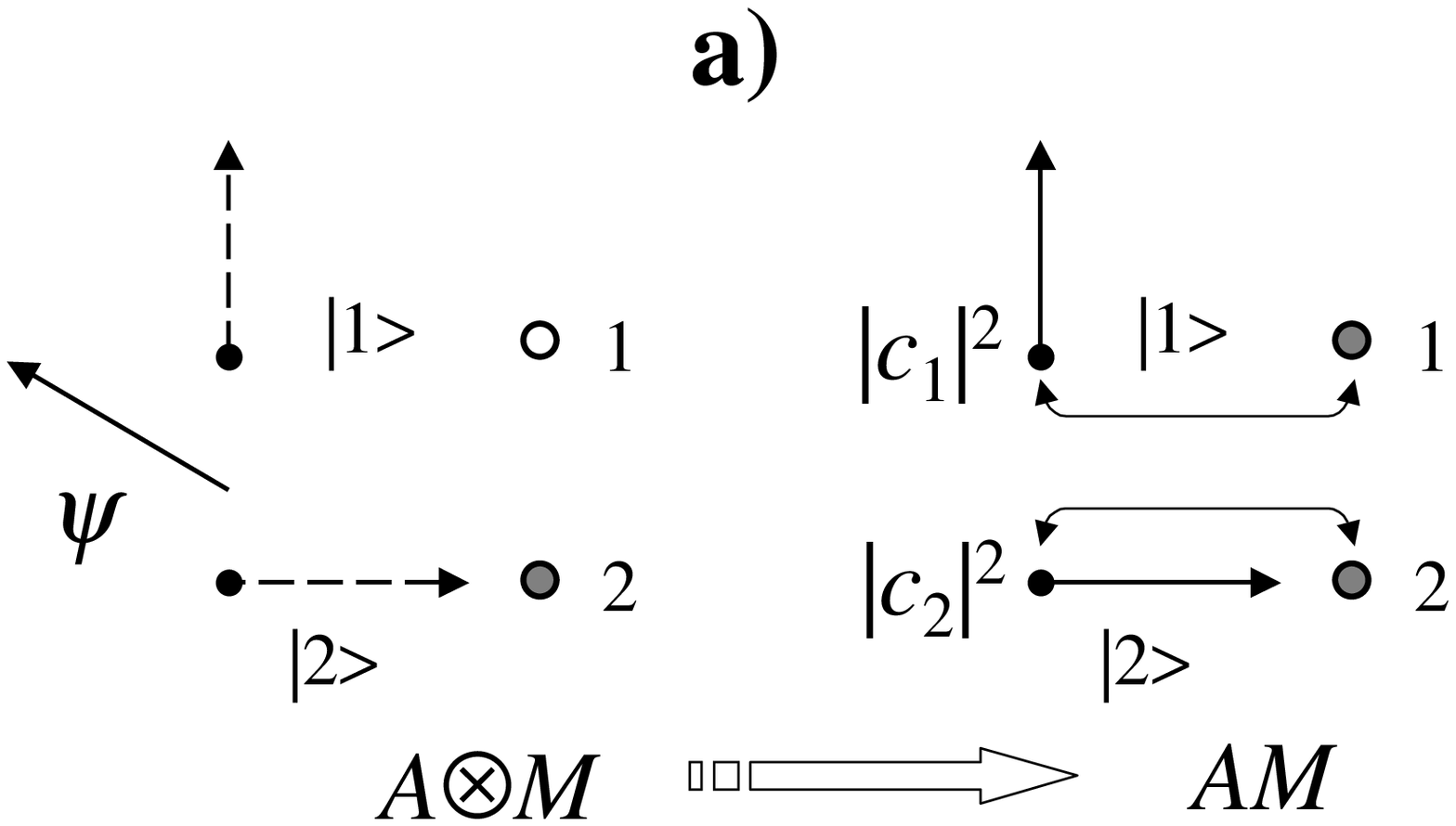}
 \hspace{0.1\textwidth}
 \epsfxsize=0.25\textwidth \epsfclipon
 \leavevmode \epsffile{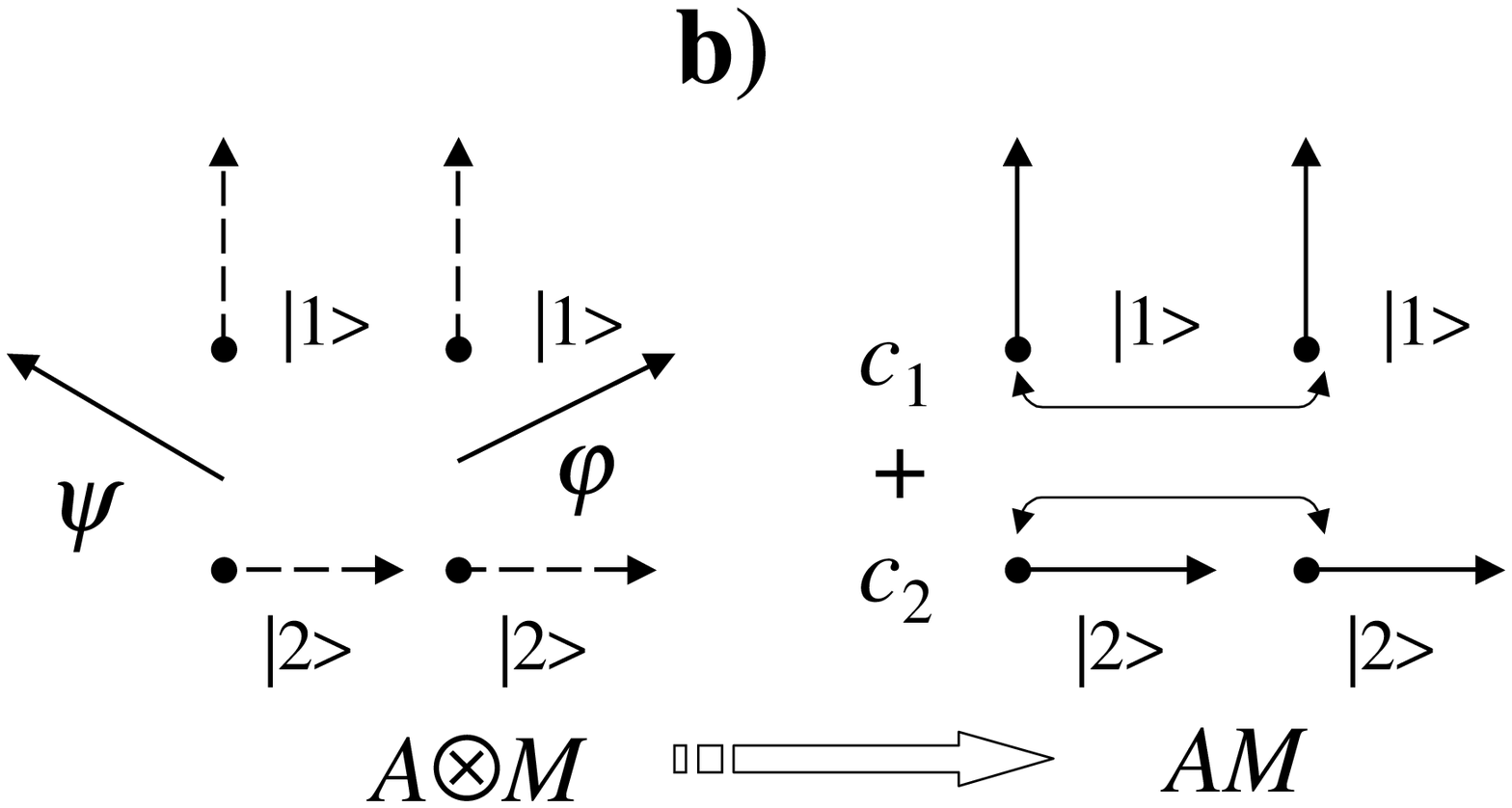}
\end{center}
\caption{Special cases of a quantum measurement. a) Standard
measurement in the bipartite setting ``quantum object--classical
pointer": the wave function $\psi$ and independent state ``2" of the
classical apparatus transfer into the statistical mixture of
totally-correlated states $\ket i\otimes\,i$ of the bipartite
system $AM$ with probabilities $|c_i|^2$ $(i=1,2)$. b) An
entangling measurement in the bipartite setting ``quantum
object--quantum pointer" for the entanglement matrix
$R_{ij}\equiv1$ (Sec.~\ref{section:cases}): an independent pure
state of the bipartite system transfers into an entangled state,
which is the coherent superposition
$c_1\ket1\ket1+c_2\ket2\ket2$.} \label{fig1}
\end{figure}

The linear transformation discussed above is defined on the direct product
${\mathbb C}_{AM} = {\mathbb C}(H_A) \otimes{\mathbb
C}(\Lambda_M)$ of the quantum-mechanical operator algebra in $H_A$
and the algebra of classical functions on the set $\Lambda_M$. This
transformation yields the pointer value $\lambda$ (independent of
its initial values $\mu$) for the density matrices of the pure
state $\hat\rho^A_\lambda=\psi\psi^+$ with wave functions $\psi$
of the measurable quantum system, which lie in the eigen subspaces
$\Phi_\lambda$ corresponding to the value $\lambda$ of the
measurable variable, i.e., $\hat
P_\lambda^A\psi=\psi\in\Phi_\lambda$. If the measurement is a
complete one, i.e., $\Phi_\lambda$ are one-dimensional, then the
transformation (\ref{rlm}) for an arbitrary density matrix
$\hat\rho^A$ in a joint initial state of the form $\hat\rho(\mu)
=\hat\rho^A\otimes p(\mu)$ describes the output mixture of
the corresponding pure states $\hat P_\lambda^A =
\ket\lambda\bra\lambda$ with probability weights
$p(\lambda)=\bra\lambda\hat\rho\ket\lambda$. A scheme clarifying
the measurement transformation structure in the described
quantum-classical system is shown in Fig.~\ref{fig1} .

The problem of the physical realizability of a quantum system
measurement procedure with the help of a classical system has attracted
considerable interest in the literature, but some principal questions
related to this problem are still under discussion
\cite{Peres,Marsh,Zhang,Sokol}. A simple example clarifying the key
mechanisms for realizing the measurement procedure (\ref{Slm})
in a closed physical system described quantum-mechanically
to preserve the quasiclassical character of the apparatus' variable, is
given in Ref. \onlinecite{QSP}.

Along with the maximally simplified description of the
measurement procedure in the form of superoperator (\ref{Slm}),
which takes into account the physical structure of the apparatus only
in the form of the classical variable $\mu\to\lambda$, a more detailed
description must at least include the quantum-mechanical variables of
the apparatus that are complimentary to the classical variable. In
this case, a minimal extension of the model leads to the replacement
of the classical apparatus' pointer with a quantum one, which
lives in the space $H_M$ with dimension $D$ equal to the number of
values of the measurable variable.

Accordingly, superoperator (\ref{Slm}) is replaced with the fully
quantum superoperator of the \emph{standard} measurement
\begin{equation}\label{MAM}
{\cal M}_{AM}=\sum\limits_\lambda\bigl(\hat P_\lambda^M {\rm
Tr}_M^{}\odot\bigr)\otimes\bigl(\hat P_\lambda^A \odot\hat
P_\lambda^A\bigr).
\end{equation}

\noindent Here the $\hat P_\lambda^A$ are the same as in Eq.\
(\ref{Slm}) and the $\hat P_\lambda^M$ are the one-dimensional projectors
corresponding to the values $\lambda$ of the apparatus' variable
$\hat M= \sum\lambda\hat P^M_\lambda$.

Projecting the density matrix of the measurable quantum system
described by the operators $\hat P_\lambda^A$ leads to the
transformation of the system's state into an incoherent
superposition of respected states with explicitly determined
values $\lambda$ of the variable $\hat A$. Operation ${\rm
Tr}_M\odot$ reflects the independence of the final state of the
apparatus from its initial state and the projectors $\hat P^M_\lambda$
describing the resulting quantum state of the pointer after the
measurement, which correspond to the measured values $\lambda$. In
the general case, the projectors $\hat P_\lambda^M$ considered
in the real physical space of the apparatus are multidimensional;
this corresponds to macroscopic systems with numerous internal
degrees of freedom of the apparatus. However, if
these internal degrees of freedom do not affect essentially
the interaction of the apparatus with the measurable quantum system,
the measurement can be adequately described in the minimal
Hilbert space $H_M$.

Superoperator (\ref{MAM}), being physically realizable, is
completely positive \cite{kraus} and, additionally, is hermitian
with respect to the scalar product $(\hat\rho_1, \hat\rho_2)={\rm
Tr}\,\hat\rho_1^+\hat\rho_2$. Then, its particular property of
idempotency, i.e., ${\cal M}_{AM}^2= {\cal M}_{AM}$, identifies an
orthogonal projector onto the subspace of unperturbed states of
the bipartite system ``object--pointer''.

In the description of the measurement procedure given above,
the quantum nature of the apparatus is not essential, though it is
virtually present in its mathematical
description. Only the pointer variable is used and the off-diagonal
quantum operators are simply not considered. Mathematically, it
means that we consider a reduced algebra of quantum events
${\mathbb B}$ of the quantum probabilistic space $(H,{\mathbb B},
{\cal P})$, where $H$ is the Hilbert space of all states allowed
in the quantum system, $\mathbb B$ is the algebra of its subspaces
$\Omega\subseteq H$ identified by the orthoprojectors
$\hat P(\Omega)$, and ${\cal P}(\Omega) ={\rm Tr}\,\hat
P(\Omega)\hat\rho$ is the quantum probability distribution defined
with the help of the density matrix. In this quasiclassical case,
the algebra of orthogonal subspaces in $H_M$ built on the eigen
subspaces $\Phi_\lambda$ of the eigen projectors $\hat
P_\lambda^M$ of the apparatus' variable $\hat M$ is used for
the description of the pointer variable ${\mathbb B}$. However,
besides the variables commuting with the pointer variable $\hat M$
there are also off-diagonal variables of the form $\hat
N=\sum_{\lambda\mu} N_{\lambda\mu} \ket{\lambda}^M \bra{\mu}^M$
that do not commute with $\hat M$ and can potentially lead to the
essentially quantum nature of the apparatus even at macroscopic
level.

The fact that in a real macroscopic system there exist variables, which
do not commute with $\hat P_\lambda^M$, is not a paradox.
In physics, pithy examples of such quantum variables in
quasiclassical systems arise, for instance, when considering
polyatomic molecules (an effective two-level model of molecular
chirality related to the chiral degree of freedom of a chiral
molecule having stable enantiomers can serve as such a pithy
example \cite{bychkov,JRS02}). Similar subsystems can be extracted
in the mathematical description of a complete set  of potentially
possible states of any macroscopic system. Typically, their
quantum nature is not essential because of the small values of
energy quanta corresponding to the transitions between discrete
energy levels.

On the contrary, there is also a number of macroscopical quantum systems,
i.e., quantum dots, superconducting Josephson junctions, and others, which
are considered as embodies of qubits in quantum information processing
\cite{QC} where the quantum nature of the apparatus can be essential. Specific
models of apparatus can, obviously, limit both the quality and fidelity of
reproduction of the measurement transformation (\ref{MAM}) due to the
macroscopic nature of the apparatus. However, as follows from the analysis
of numerous specific models in the literature, such
limitations do not forbid realizations of such measurement models.

An initial density matrix of the form $\hat\rho^A{\otimes}\hat\rho^M$
is transformed by the superoperator (\ref{MAM}) into the density matrix
\begin{equation}\label{rAM}
\hat\rho^{AM}=\sum\limits_\lambda\bigl(\hat P_\lambda^A\hat\rho^A\hat
P_\lambda^A \bigr)\otimes\hat P_\lambda^M
\end{equation}

\noindent characterizing the state with the coinciding variables
``$\hat A{=}\hat M$'', i.e., in strict form $(\hat A-\hat
M)\hat\rho^{AM}=0$. This state is an \emph{incoherent} statistical
mixture of quantum states characterizing each of the variables.
The transfer of information between the quantum system $A$ and
apparatus $M$ is realized in such a state through the classical
variable $\lambda$. Quantum fluctuations therefore exist only
virtually as uncertainty in the physical variables, which do
not commute with $\hat A$ and $\hat M$ and are therefore not
determined explicitly.

The above-described semiclassical concept of the measurement is not a general
one \cite{Neumann55,Ludwig,Stenholm}. Moreover, generation of
entanglement by means of quantum measurement has been a topical issue for the last decade
\cite{Kuzmich,Duan,Kozhekin,Molmer,Jakob}. Therefore, a generalization of
the standard measurement approach can essentially extend the concept of
quantum measurement.

For example, an essentially quantum isometric (at fixed $\psi_M$)
transformation $\psi_A\otimes \psi_M \to \sum{\braket{k} {\psi_A}}_A
{\ket k}_A{\ket k}_M$ of an arbitrary pure state into an entangled
state can be interpreted as a measurement of the variable $\hat A=\sum
\lambda_k{\ket k}_A{\bra k}_A$ with the help of the apparatus' variable
$\hat M= \sum\lambda_k{\ket k}_M{\bra k}_M$. The corresponding
generalization of the measurement superoperator (\ref{MAM}) has the
form
\begin{equation}\label{MAMp}
{\cal M}_0=\sum\limits_{kl}\bigl(\hat P_{kl}^M{\rm Tr}_M^{}
\odot\bigr)\otimes\bigl(\hat P_{kk}^A\odot\hat P_{ll}^A\bigr),
\end{equation}

\noindent where the $\hat P_{kl}=\ket k\bra l$ are the one-dimensional
projectors from $\ket l$ onto $\ket k$ in the respective Hilbert spaces $H_A$ and
$H_M$. For a pure state $\hat\rho^A=\ket{\psi_A} \bra{\psi_A}$, this transformation
ensures the resulting pure entangled state of the composite
system. Note, however, that this does not mean the absence of
dequantization of the initial state, because the latter is represented in
the final state only by the diagonal orthoprojectors $\hat P_{kk}^A$ and
$\hat P_{ll}^A$.

The above formulas for the measurement superoperators ${\cal M}_{AM}$
and ${\cal M}_0$ can be generalized to an intermediate representation of
the form
\begin{equation}\label{Mgen}
{\cal M}=\sum\limits_{kl}R_{kl}\bigl(\hat P_{kl}^M{\rm Tr}_M^{} \odot
\bigr)\otimes\bigl(\hat P_{kk}^A\odot\hat P_{ll}^A\bigr),
\end{equation}

\noindent which describes an \emph{entangling measurement} with the Hermitian
\emph{entanglement matrix} $R_{kl}$. This matrix is chosen to ensure the
complete positivity and normalization condition for the measurement
transformation. To our knowledge, Eq.\ (\ref{Mgen}) is the most
general form of the entangling measurement superoperator based on the linear
combination of the input-output projectors, which is compatible with the ideal
measurements concept.

At $R_{kl}=\delta_{kl}$, Eq.\ (\ref{Mgen}) simplifies to the standard
quantum measurement (\ref{MAM}), if one identifies the projector sets
with corresponding indices $k$ and $\lambda$. The projectors $\hat P^A_{kk}$
in Eq.\ (\ref{Mgen}) characterize information in the quantum system to be
measured and its statistical properties are determined by the density
matrix $\hat\rho^A$. Information that is finally measured by the apparatus
is represented by the projectors $\hat P^M_{kl}$, and its statistical
properties are determined only by the density matrix $\hat\rho^A$ and are
invariant with respect to the choice of a basis in the state space $H_M$ of the
apparatus.

The measurement transformation (\ref{Mgen}) creates an entanglement in the
bipartite system ``quantum object system--apparatus" whose value may
vary between zero and its maximum value (which depends on $R_{kl}$).

Taking the basis set of the density matrices in the form
$\hat\rho^A{\otimes} \hat\rho^M$, we obtain for the transformation of
the joint density matrix:
\begin{equation}\label{rhoAM}
\hat\rho^{AM}={\cal M}\,\hat\rho^A\otimes\hat\rho^M = \sum R_{kl}
\rho^A_{kl}\hat P_{kl}^A\otimes\hat P_{kl}^M.
\end{equation}

\noindent For the initial density matrix of the composite system of
general form $\hat\rho^{AB}$, the resulting matrix has the same
form (\ref{rhoAM}), keeping in mind that the matrix elements
$\rho^A_{kl}$ correspond to the partial density matrix
$\hat\rho^A={\rm Tr}_B \hat\rho^{AB}$. Therefore, the density
matrix of a composite system is formed in the process of
measurement as a result of the dubbing ${\ket
k}^A\to\ket{k}^A\ket{k}^A$ of the chosen object basis, accompanied
by multiplication of the corresponding elements of the initial
density matrix of the object by those of the entanglement
matrix.

\section{\label{section:cases}Special cases}

Let us assume that the initial state of a quantum system is a pure
state, i.e., $\hat\rho^A=\psi_A \psi^+_A$ with
$\psi_A=(c_1,\dots,c_D)$, and the measurement procedure is
complete, which means that all the projectors $\hat P_{kl}$ are
one-dimensional. Then, it follows from Eq.\ (\ref{rhoAM}) that
\begin{equation}\label{rhoAMs}
\hat\rho^{AM}=\sum\limits_k\varkappa_k\ket{\ket k}\bra{\bra
k},\quad \ket{\ket k}=\sum\limits_i e_{ki} \ket i\ket i,
\end{equation}

\noindent where the $\varkappa_k$, $e_k=(e_{k1},\dots,e_{kD})$ are the
eigenvalues and the respective normalized eigenvectors of the matrix
$\tilde\rho_{kl}= R_{kl}\rho_{kl}=R_{kl} c_k c^*_l$. This means that the
pure state is transformed into an incoherent mixture of pure entangled
states of the bipartite system ``object--pointer'', which are orthogonal to
each other and have different degrees of entanglement that depends on
the entanglement and density matrices. For the case of $\hat\rho^A=
\ket{k_0}\bra{k_0}$, we have $c_k=\delta_{kk_0}$ and the corresponding joint
density matrix $\hat\rho^{AM}=\ket{k_0}^A\bra{k_0}^A \otimes \ket{k_0}^M
\bra{k_0}^M$, which coincides with the density matrix resulting from the standard
measurement ${\cal M}_{AM}$ of the observable $\hat A=\sum\lambda_k
\ket{k}^A\bra{k}^A$. For the case of pure {\em maximum uncertainty}
state $\hat\rho^A=(1/D)\sum\ket{k}\bra{k}$, we have
$\hat\rho^{AM}=\sum_{kl}(R_{kl}/D)\ket{\ket k}\bra{\bra l}$ with
$\ket{\ket k}=\ket k \ket k$, where $R_{kl}/D$ is the object--pointer density
matrix in the basis of the dubbed (``cloned'') states.

For the entanglement matrix $R_{ij}=\delta_{ij}$, all the
eigenvalues $\varkappa_k=|c_k|^2$ in Eq.\ (\ref{rhoAMs}) are
represented by the eigenvectors $e_{ki}=\delta_{ki}$ and the
degree of entanglement of each of the $\ket{\ket k}$ states is equal to
zero. Thus, a completely incoherent mixture of states of the
measurable variable $\hat A=\sum\lambda_i\ket i\bra i$ with
determined $\lambda_i$ is formed.

For the entanglement matrix $R_{ij}\equiv1$, which has the only
nonzero eigenvalue $\varkappa_k=1$ and respective eigenvector
$e_{ki}=c_i$, Eq.\ (\ref{rhoAMs}) gives us a pure state, which
is represented by the single vector
\begin{displaymath}
\ket{\ket k}= \sum c_i \ket i\ket i.
\end{displaymath}

\noindent The state $\ket{\ket k}=\sum c_i \ket i \ket i$,
corresponding to the state in Eq.\ (\ref{rhoAMs}), coincides with
the state formed after the \emph{quantum duplication}
transformation \cite{PRA00}. The degree of its entanglement $E$
can be estimated as the entropy $S[P]$ of the probability
distribution $P(i)=|c_i|^2$ of all possible values $\lambda_i$ of
the measurable variable. We can \emph{always} receive a maximum
possible degree of entanglement $E=\log_2D$ by choosing the
measurable variable $\hat A$ as having maximum uncertainty in
the state $\psi$, which corresponds to the vector representation
$c_i\equiv1/\sqrt{D}$ and the uniform distribution $\psi\to P(i)=1/D$.
Therefore, for the maximally entangling measurement, vacuum quantum
fluctuations of the measurable variable are transferred into the
corresponding entanglement of the bipartite system
``object--pointer''.

For the general case of a mixed initial state $\hat\rho^A=\sum\rho_l
\ket l\bra l$, an incoherent mixture of $D$ orthogonal (at fixed
$l$) sets of entangled states $\ket{\ket{k,l}}$ is formed. These
sets, however, are not necessarily orthogonal at different values
of $l$, but this nonorthogonality is largely of formal character
and physically does not mean nonorthogonality of the states of one
and the same Hilbert spaces. When one considers an incoherent
mixture of states, phase uncertainty is due to an additional
degree of freedom, and incoherence means that we consider
physically \emph{distinguishable} (orthogonal \cite{RTE02,fuchs})
states. In fact, the orthogonal wave functions $\psi,\varphi$
correspond to two physically distinguishable states $i,j$ that
describe different phases in the composite states
$\psi=\ket\alpha\ket i$, $\varphi=\ket\beta\ket j$ even for
nonorthogonal states $\alpha$ and $\beta$. Keeping this in mind, in
a more detailed quantum description of a composite system, which
includes all physically valuable degrees of freedom, quantum
operators of the corresponding physical subsystems do commute.

\section{\label{section:matrix}General structure of the quantum measurement superoperator and
analysis of a two-dimensional model}

Let us first define the constraints imposed on the quantum measurement
superoperator by the normalization condition and positivity property. From
Eq.\ (\ref{rhoAM}) we obtain ${\rm Tr}\,\hat\rho^{AM}=\sum
R_{kk}\rho^A_{kk}$, which, with an arbitrary choice of
$\hat\rho^A$, immediately gives  the normalization condition $R_{kk}\equiv1$. The
positivity requirement and, simultaneously, the complete positivity properties
require the positivity of the matrix $\hat\rho^A_e=
\left(R_{kl}\rho^A_{kl}\right)$ for an arbitrary positive matrix $\hat
\rho^A= \left(\rho^A_{kl}\right)$. Then, using the spectral representation of
both these matrices, one can readily show that a necessary and sufficient
condition for this requirement is the positivity of the entanglement matrix
$R=\left(R_{kl}\right)$.

A repeated entangling measurement, i.e., the repeated application of the same
measurement superoperator on the same apparatus--system Hilbert space,
leads, on account of the measurement superoperator (\ref{Mgen}), equality
$R_{kk}\equiv1$, and vanishing of the off-diagonal elements of the pointer
density matrix after tracing out ${\rm Tr}_M\odot$, to the relation
\begin{equation}\label{MM}
{\cal M}^2={\cal M}_{AM}.
\end{equation}

\noindent Therefore, a repeated entangling measurement leads to
entanglement destruction and the resulting transformation is equal
to the standard measurement. This entanglement destruction in the initial
state of the ``quantum object--pointer'' is due to a ``resetting" of the apparatus
needed to achieve independence of the final system--apparatus state from
the initial apparatus state.

Eq.\ (\ref{MM}) is valid for any entangling matrix $R$, i.e., the
entangling measurement ${\cal M}$ is the ambiguous square root of the
standard measurement ${\cal M}_{AM}$, which reveals in the spectrum
structure of the corresponding matrices representing ${\cal M}$.

As an example, let us consider a two-level system with ${\rm
dim}\,H_A=2$ for which the positivity criterion gives the
following general form for the entanglement matrix:
\begin{equation}\label{Rq}
R=\left(
{\begin{array}{cc}
  1   & q \\
  q^* & 1
\end{array}}\right),\quad |q|^2\le1.
\end{equation}

The eigenvalue equation for the measurement superoperator
\begin{equation}\label{Mrho}
{\cal M}\,\hat\rho^{AM}=\lambda\,\hat\rho^{AM}
\end{equation}

\noindent can be solved analytically once we have expressed it in
the form
\begin{widetext}
\begin{displaymath}
\sum_{kl}\sum_m R_{kl}\rho^{AM}_{klmm}\hat P_{kl}^A\otimes\hat
P_{kl}^M=\lambda\sum_{kl}\sum_{mn}\rho_{klmn}^{AM}\hat P_{kl}^A
\otimes\hat P_{mn}^M\to
R_{kl}\Bigl(\sum_\mu\rho^{AM}_{kl\mu\mu}
\Bigr)\delta_{km}\delta_{ln}=\lambda\rho_{klmn}^{AM}.
\end{displaymath}
\end{widetext}

\noindent For $D=2$, the dimension of the problem is limited to 16
possible values of the four-dimensional index $klmn$ of the ``density
matrix" (in the eigenvalue problem, in addition to  physically valuable
density matrices, arbitrary operators are considered, as well).

On account of the vanishing off-diagonal elements in the
transformed density matrix $\rho^{AM}_{klmn}$ on the left-side of
the above equation, we find that eight right eigen null-vectors
$\hat{e}_{AM}^{0k}$, $k=1,\dots,8$ corresponding to the eigenvalue
$\lambda=0$ are described by the following operators:
\begin{equation}\label{rAM0k1}
\hat{e}_{AM}^{0k}=\left\{\begin{array}{c}\!\hat\rho_A^{0k}\otimes\hat
P_{12}^M,\;k=1,2,3,4,\\
\hat\rho_A^{0k}\otimes\hat
P_{21}^M,\;k=5,6,7,8,
\end{array}\right.
\end{equation}

\noindent which have zero diagonal matrix elements
$\rho^{AM}_{klmm}$ along $M$ with arbitrary density matrices
$\hat\rho_A^{0k}$. Freedom in choosing them is due to the
eightfold degeneracy and related to the arbitrary choice of four
linearly independent basis vectors (\ref{rAM0k1}) corresponding to
the related basis operators $\hat P_{12}^M$, $\hat P_{21}^M$ of
the pointer $M$.

The other four zero eigenvectors satisfy the relation $\hat
\rho_{kl11}^{AM}=-\hat\rho_{kl22}^{AM}$ and have the form
\begin{equation}\label{rAM0k2}
\hat{e}_{AM}^{0k}\,=\hat\rho_A^{0k}\otimes\bigl(\hat P_{22}^M- \hat
P_{11}^M \bigr),\quad k=9,10,11,12
\end{equation}

\noindent with arbitrary linearly independent density matrices $\hat
\rho_A^{0k}$.

Finally, for two nonzero eigenvectors with $\lambda=1$ we have a pair
of linearly independent functions $\rho_k=\delta_{k1}, \delta_{k2}$ satisfying
the relations $\rho^{1,k}_{klmn}=\rho_k\delta_{kl} \delta_{km}
\delta_{ln}$ and corresponding operators
\begin{equation}\label{rAM1k}
\hat{e}_{AM}^{1k}\,=\left\{\begin{array}{c}\hat{P}^A_{11}\otimes
\hat{P}^M_{11}\\ \hat{P}^A_{22}\otimes \hat{P}^M_{22}
\end{array}\right.,\quad k=13,14.
\end{equation}

\noindent These two operators provide a basis of the convex set
$p\,\hat e^{1,13}_{AM}+(1-p)\hat e^{1,14}_{AM}$ ($0\le p\le 1$) of
the density matrices, which are not changed in the measurement
transformation.

The last two linearly independent operators $\hat e^{015}_{AM}=\hat
P^A_{12}\otimes\hat I^M$ and $\hat e^{016}_{AM}=\hat P^A_{21}\otimes\hat
I^M$ are the eigen operators with eigenvalue equal to zero only at $R_{kl}=
\delta_{kl}$ when the measurement is a standard measurement, \emph{not}
an entangling measurement. In the general case, the superoperator $\cal M$
lacks two eigenvectors, because it is not described by a matrix of simple
structure similar to the single-mode fermion annihilation operator $\hat
a=\left(\begin{array}{cc} 0&1\\0&0\end{array} \right)$, which has a single
non-vanishing right eigenvector $e_0=(1,0)$ \cite{note1}.
Accordingly, Eq.\ (\ref{MM}) is realized as the relation $\hat a^2=0$,
which washes out dependence of the squared operator on the entangling
parameter $q$.

The corresponding linear subspace $c_{15}\hat e^{015}_{AM}
+c_{16}\hat e^{016}_{AM}$ contains only density matrices, which
have no physical meaning. Nevertheless, this subspace cannot be
excluded from the complete 16D-space, because it is included into
the cone of all positive hermitian density matrices.

The entangling measurement superoperator matrix in the ``eigen"
basis $\hat e_{AM}^{\lambda k}$ has the form:
\begin{equation}
 \label{Mstructure}
{\cal M}=\left(
\begin{array}{ccccccc}
  1 & 0 &\hat{\underline{0}}& 0 & 0 & 0 & 0 \\
  0 & 1 &\hat{\underline{0}}& 0 & 0 & 0 & 0 \\
\hat0|&\hat0|&\hat{\rm O} &\hat0|&\hat0|&\hat0|&\hat0|\\
  0 & 0 &\hat{\underline{0}}& 0 & 0 & q & 0 \\
  0 & 0 &\hat{\underline{0}}& 0 & 0 & 0 & q^* \\
  0 & 0 &\hat{\underline{0}}& 0 & 0 & 0 & 0 \\
  0 & 0 &\hat{\underline{0}}& 0 & 0 & 0 & 0
\end{array}\right),
\end{equation}

\noindent where $\hat O$ is a $10\times10$ zero matrix, and $\hat{
\underline{0}}$ and $\hat0|$ are 10-component zero bra- and
ket-vectors, respectively. In the matrix (\ref{Mstructure}), 4th and 5th
lines correspond to the transversal--transversal basis operators and
two bottom lines correspond to the two non-eigen operators $\hat P_{12}
\otimes \hat I/D$, $\hat P_{21}\otimes\hat I/D$. The subspaces
corresponding to the matrix (\ref{Mstructure}) are 2D invariant,
12D zero, and 2D improper subspaces.

\section{\label{section:exchange}Can entangling measurement be used for entanglement transfer?}

In many applications of quantum information processing algorithms, such as
practically interesting cryptographic protocols, require
realization of a pair of spatially separated entangled quantum systems
\cite{QC,preskill}. They serve as resources for quantum information
engineering and developing technologies for their creation is of prime
importance \cite{UFN01}. Deterministic creation of spatially separated
entangled quantum systems is a difficult problem to solve. By contrast,
pairs of entangled subsystems exist naturally and spontaneously
\emph{within} many spatially-localized physical systems.

For example, conservation of the total momentum of an atom is not
related to the separate conservation of its components, orbital and spin
momenta. Thus, the eigen states of the atom are, in the general case,
entangled states related to subsystems describing orbital and spin
momenta separately. Another example is laser excitation of
ro-vibrational states in molecules, which results in the entanglement
between the vibrational and rotational degrees of freedom.

Now, the following question arises, ``Can these \emph{naturally} entangled states be used for
a transfer of their entanglement onto an entanglement of spatially
separated quantum systems with the help of an entangling measurement?".

To answer this question, let us consider four systems $A$, $B$, $M$, and $N$
with an initial state $\hat\rho^{ABMN}=\hat\rho^{AB}\otimes\hat\rho^M\otimes
\hat\rho^N$. We assume that two of them, $A$ and $B$, are in the initially entangled
state $\hat\rho^{AB}$, whereas systems $M$ and $N$ are initially independent and
spatially separated from $A$, $B$. We will then check if it is possible to
transfer entanglement from the bipartite system $A$--$B$ onto the bipartite
system $M$--$N$ by applying two independent transformations ${\cal M}_A$ and
${\cal M}_B$ to the subsystems $A$--$M$ and $B$--$N$, respectively. The corresponding
joint superoperator has the form
\begin{eqnarray*}
{\cal M}_B\otimes{\cal M}_A&=&\sum\limits_{klmn}R_{kl}R_{mn}\bigl[\hat
P^B_{mn}\otimes\bigl(\hat P^N_{mm}\odot\hat P^N_{nn}\bigr){\rm Tr}_B
\odot\bigr]\\ \nonumber
&&\otimes \bigl[\hat P^A_{kl}\otimes\bigl(\hat P^M_{kk}\odot\hat
P^M_{ll}\bigr){\rm Tr}_A \odot\bigr],
\end{eqnarray*}

\noindent where systems $M$ and $N$ are treated as pointers for the measurements
of systems $A$ and $B$. The resulting state of the bipartite system $M$--$N$ can then be obtained
by tracing out $AB$ that leads to the transformations $\hat P^B_{mn}\to \delta_{mn}$,
$\hat P^A_{kl}\to \delta_{kl}$ and, on account of $R_{kk}= R_{nn}$, can be written as
\begin{displaymath}
\hat\rho^{MN}=\sum_k\bigl(\hat P^N_{kk} \hat\rho^N\hat P^N_{kk}\bigr)\otimes\sum_n\bigl(
\hat P^M_{kk}\hat\rho^M \hat P^M_{kk}\bigr).
\end{displaymath}

\noindent This means that simultaneous measurements in the systems $A$ and $B$
always produce two \emph{uncorrelated} dequantized states of the systems $M$ and $N$
if any quantum correlations with other systems are neglected. Therefore, an entangling
measurement cannot be used for entanglement transfer onto spatially separated systems.

\section{\label{section:entangl}Quantitative characteristic of entanglement due to
entangling measurement}

In accordance with Secs \ref{section:definitions},\ref{section:cases}, the entangling
measurement superoperator creates entanglement in the
bipartite system ``object--pointer'', which does not depend on the initial
state of the pointer and is defined only by the entanglement matrix and the
initial state of the quantum system in the eigen basis of the measurable
variable (or a set of commuting variables). Generally, two types of
created entanglement due to entangling measurement
described by the respective coherent information (i.e., \emph{preserved
entanglement} \cite{Barnum01,SPIE2001a}) are of interest to us:
one-time entanglement describing one-time states $\hat\rho_{AM}$
of the system $A$ and the apparatus $M$, and two-time entanglement
describing how the initial state of the system $A$ is linked to the resulting
state of the apparatus $M$ in terms of the initial density matrix $\hat\rho_A$
and the superoperator ${\cal N}$ of a two-time channel $A\to M$ (see Eq.\
(\ref{2te})) \cite{PRA00}.

In the first case, for the one-time channel $A\rightleftarrows M$
we have the \emph{entanglement measure} $E=S[\hat\rho^M]- S[\hat
\rho^{AM}]$  (which is equal to $S[\hat\rho^A]- S[\hat
\rho^{AM}]$) with $\hat\rho^{AM}$ given by Eq.\ (\ref{rhoAM}).
Expressing the entropy via the matrix elements
$R_{kl}\rho_{kl}$, on account of $R_{kk}\equiv1$, we obtain
\begin{equation}\label{ME}
E=S[(\rho_{kk})]-S[(R_{kl}\rho_{kl})],
\end{equation}

\noindent where the entropies $S$ are calculated for the diagonalized
and complete density matrices $\tilde \rho_{kl}=R_{kl}\rho_{kl}$,
respectively.

The degree of entanglement created after the entangling
measurement defined by Eq.\ (\ref{ME}) is always positive by
contrast with the coherent information, which can be negative for an
arbitrary channel. Such induced entanglement vanishes for
diagonal density matrices, which means that coherence between the
measured states, which is transferred after the measurement onto
the pointer, is absent before the measurement.

For the case of a pure state with maximum indeterminateness of the
measurable variable, i.e. $\rho_{kl}\equiv 1/D$, an induced entanglement due to
the entangling measurement has the form
\begin{displaymath}
E=\log_2D+\sum r_k\log_2 r_k,
\end{displaymath}

\noindent where $0\le r_k\le1$ are the eigenvalues of the normalized
entanglement matrix $R_{kl}/D$. For the maximum coherency, i.e. for
$R_{kl}\equiv1$, we obtain the maximum possible value $E=\log_2D$ of
the entanglement due to the entangling measurement.

The superoperator of the two-time channel $A\to M$, i.e., a channel that links
the initial state of the quantum system and final state of the apparatus,
has the following form \cite{PRA00}:
\begin{equation}\label{2te}
{\cal N}={\rm Tr}_A{\cal M}\bigl(\odot\otimes\hat\rho^M\bigr),
\end{equation}

\noindent where the substitution symbol ``$\odot$" describes dependence on
the initial state of the quantum system $\hat\rho^A$. The channel defined this
way allows one
to use the original definition of the coherent information \cite{barnum98}.

On account of the measurement superoperator
structure (\ref{Mgen}), the superoperator (\ref{2te})  does not depend on the initial pointer's state
$\hat\rho^M$. Thus, with the help of the
transformation ${\cal M}$ we can readily conclude that tracing out the initial state leads
to the diagonalization of the output density matrix $\rho_{kk}^M =
\rho_{kk}^A$ and its dependence solely on the diagonal part of the initial
density matrix of the measurable quantum system. Such a transformation for
the coherent information $S[\hat\rho^M]-S[({\cal N}\otimes{\cal
I})\Psi_{AR}\Psi_{AR}^+]$, where $\Psi_{AR}$ defines the initial state of
the input and the reference system $R$ corresponding to the density
matrix at the input $\hat\rho^A$ always yields a zero value due to the
fact that both density matrices are diagonal and their diagonal elements
are equivalent.

Therefore, entanglement after the measurement is created only for
one-time states, but two-time entanglement does not exist because of
the destruction of initial coherency.

\section{\label{section:conculsions}Conclusions}

In conclusion, we have studied a natural mathematical
generalization of the standard quantum measurement on the
entangling quantum measurement, which creates an entanglement between the
measurable quantum system and the apparatus in the bipartite
setting ``quantum object--pointer''. The entangling
measurement procedure is defined, as well as the standard measurement
procedure, by the choice of measurable variables and,
additionally, by the entanglement matrix. Such a procedure can be
physically realized with the help of an apparatus that can have
either microscopic or macroscopic nature. In the latter case, we
deal with an ``ideal" measurement transformation.

Repeated entangling measurement results in the
standard incoherent measurement transformation. Thus, the
entangling measurement superoperator can be represented by the
ambiguously determined square root of the standard measurement
superoperator. This ambiguity, as has been illustrated in the
two-dimensional example, is due to the incompleteness of the corresponding
superoperator's eigenvector system.

The entangling measurement, as we have proved, cannot be used for
entanglement transfer from a bipartite system to another one.

It has also been shown that the entangling measurement creates a
one-time entanglement in the bipartite system ``quantum object--pointer''
whose degree depends on the entanglement matrix and the initial state of the
quantum system and is bounded above by the logarithm of the number of
measurable values. The degree of two-time entanglement is always equal to
zero due to the complete decay of initial coherency. This means that
only dequantized information about the initial state of the quantum
system is preserved.

It is argued that the coherent information is a valid tool for
characterizing entanglement created in the bipartite system
``quantum object--pointer''.

\begin{acknowledgments}
This work was supported in part by the Russian Foundation for Basic Research
under grants Nos. 01--02--16311, 02--03--32200, and by INTAS grant INFO 00--479.
\end{acknowledgments}



\begin{references}
\bibitem{note1}
Due to this property of the entangling measurement, superoperator
relation (\ref{MM}) is valid. Without this property, it follows
from Eq.\ (\ref{MM}) that the superoperator can be represented in
the form of the positively defined square root of the standard
measurement superoperator, which does not depend on the
entanglement matrix $R_{kl}$ characterizing the entangling
measurement superoperator. For $R_{kl}\equiv\delta_{kl}$, this
property is missed and, therefore, the
``square-root-representation" is valid.
\end{references}
\end{document}